\documentstyle[prl,aps,floats,epsf]{revtex}
\begin{document}

\baselineskip=12pt

\twocolumn[\hsize\textwidth\columnwidth\hsize\csname@twocolumnfalse\endcsname

\title
{Level Crossing Analysis of Burgers Equation in $1+1$ Dimensions }

\author
{M. Sadegh Movahed $^{a,b}$, A. Bahraminasab,$^{a,c}$, H. Rezazadeh
$^d$, A. A. Masoudi $^{d,e}$ }
\address
{\it $^a$ Dept. of Physics , Sharif University of Technology,
P.O.Box 11365-9161, Tehran, Iran.\\
$^b$ Institute for Studies in Theoretical Physics and Mathematics,
P.O.Box 19395-5531, Tehran, Iran \\
%$^c$  Iran Space Agency,  PO Box  199799-4313, Tehran, Iran\\
 $^c$
International Center for Theoretical Physics, Strada Costiera
11, I-34100 Trieste, Italy \\
$^d$ Dep. of Physics, Islamic Azad university, Branch of North
Tehran, P.O.Box 19585-936, Tehran, Iran\\
$^e$  Department of Physics, Alzahra University,P.O.Box 19834,
Tehran, Iran
 }
\maketitle

\begin{abstract}

We investigate the average frequency of positive slope
$\nu_{\alpha}^{+}$, crossing the velocity field  $u(x)- \bar u =
\alpha$ in the Burgers equation. The level crossing analysis in
the inviscid limit and total number of positive crossing of
velocity field  before creation of singularities are given. The
main goal of this paper is to show that this quantity,
$\nu_{\alpha}^{+}$, is a good measure for the
fluctuations of velocity fields in the Burgers turbulence.      
\end{abstract}
\hspace{.3in}
\newpage]

\section{Introduction}
The Burgers equation is the simplest nonlinear generalization of
the diffusion equation. The N-dimensional forced Burgers equation
 \begin{eqnarray}
\partial_t\vec{u}+(\vec{u}.\vec{\nabla})\vec{u} =\nu\nabla^2\vec{u}+\vec f(\vec{x}, t)\hspace{1cm}
\end {eqnarray}
which describes the dynamics of a stirred, pressure less and
vorticity-free fluid, has found interesting applications in a wide
range of non-equilibrium statistical physics problems. It arises,
for instance, in cosmology where it is known as the adhesion model
\cite{cos}, in vehicular traffic \cite{traf} or in the study of
directed polymers in random media \cite{med}. In the Burgers
equation  if the velocity field  is a gradient field
$\vec{u}(\vec{x},t)=-\vec{\nabla}\psi(\vec{x},t)$ and the random
force is a gradient random force
$\vec{f}(\vec{x},t)=-\vec{\nabla}F(\vec{x},t)$, then
 associated Hamilton –
Jacobi equation, satisfies in the following equation as:
 \begin{eqnarray}
\partial_t\psi =\nu\nabla^2\psi+\frac{1}{2}(\vec{\nabla}\psi)^2+ F(\vec{x}, t)
\end {eqnarray}
where $\nu$ is the viscosity, recently it has been frequently
studied as a nonlinear model for the motion of an interface under
deposition \cite{KPZ}. The case with large-scale forcing was
considered in Refs. \cite{ya,ch} as a natural way to pump energy
in order to maintain a statistical steady state. Burgers equation
is then a simple model for studying the influence of
well-understood structures (shocks, preshocks, etc) on the
statistical properties of the flow. As it is well known, eq.(1)
in the limit of vanishing viscosity ($\nu\to 0$) displays after a
finite time dissipative singularities, namely shocks,
corresponding to discontinuities in the velocity field. In the
presence of large-scale forcing, it was recently stressed for the
one-dimensional case \cite{w1,w2} and also for higher dimensions
\cite{r,j,bah05}, that the global topological structure of such
singularities is strongly related to the boundary conditions
associated to the equation. More precisely, when, for instance,
space periodicity is assumed, a generic topological shock
structure can be outlined. It plays an essential role in
understanding the qualitative features of the statistically
stationary regime. So far, the singular structure of the forced
Burgers equation was mostly investigated in the case of
finite-size systems with periodic boundary conditions. It is
however frequently of physical interest to investigate instances
where the size of the domain is much larger than the scale, so as
to examine, for example, the role of
Galilean invariance \cite{bara}. \\
Here we describe the level crossing analysis in the context of
vorticity - free fluid. In the level crossing analysis we are
interested in determining the average frequency ( in spatial
dimension ) of observing of the definite value for velocity
function $u(x)-\bar{u}=\alpha $ in fluid, $\nu _\alpha ^{+}$, from
which one can find the averaged number of crossing the given
velocity in sample with size L. The average number of visiting the
velocity $u(x)-\bar{u}=\alpha $ with positive slope will be
$N_\alpha ^{+}=\nu _\alpha ^{+}L$. It can be shown that the $\nu
_\alpha ^{+}$ can be written in terms of joint probability
distribution function (PDF) of $u(x)-\bar{u}$ and its gradient.
Therefore the quantity $\nu _\alpha ^{+}$ carry the whole
information of fluid which lies in joint PDF of velocity and its
gradient fluctuations. This work aims to study the frequency of
positive slope crossing (i.e. $\nu _\alpha ^{+}$) in time $t$ on
the vorticity - free fluid in a sample with size L.
 We describe a quantity $N_{tot}^{+}$ which is defined as $
N_{tot}^{+}=\int_{-\infty }^{+\infty }\nu _\alpha ^{+}d\alpha $ to
measure the total number of crossing the velocity of fluid with
positive slope.
 The $N_{tot}^{+}$  and the path which is constructed velocity of fluid are in the
same order. It is expected that in the stationary state the
$N_{tot}^{+}$ to become size dependent. Although we  exactly
determine the velocity dependence of $\nu _\alpha ^{+}$ for
Burgers equation in the inviscid limit and before creation of
singularities, we compute the time dependence of $N_{tot}^{+}$ ($\nu _\alpha ^{+}$) numerically.\\
 This paper is organized as follows: In
section II we discuss the connection between $\nu _\alpha ^+$ and
underlying probability distribution functions (PDF) of a fluid
\cite{ris}. In section III we derive the integral representation of
$\nu _\alpha ^+$ for the Burgers equation in 1+1 dimensions and in
the inviscid limit before the creation of singularities. Section IV
closes with a discussion of the present results.

\section{ The Level Crossing Analysis of  stochastic Processes }

Consider a sample function of an ensemble of functions which make
up the homogeneous random process $u(x,t)$. Let $n_{\alpha}^{+}$
denote the number of positive slope crossing of $u(x)- \bar u =
\alpha$ in time $t$ for a typical sample size $L$ (see fig.(1) )
and let the mean value for all the samples be $N_{\alpha}^{+}(L)$
where:
\begin{equation}
N_{\alpha}^{+}(L)=E[n_{\alpha}^{+}(L)].
\end{equation}

Since the process is homogeneous, if we take a second interval of
$L$ immediately following the first we shall obtain the same
result, and for the two intervals together we shall therefore
obtain:
\begin{equation}
N_{\alpha}^{+}(2L)=2N_{\alpha}^{+}(L),
\end{equation}
from which it follows that, for a homogeneous process, the
average number of crossing is proportional to the space interval
$L$. Hence:
\begin{equation}\label{2}
N_{\alpha}^{+}(L)\propto L,
\end{equation}
or:
\begin{equation}\label{3}
N_{\alpha}^{+}(L)=\nu^{+}_{\alpha} L.
\end{equation}
which $\nu_{\alpha}^{+}$ is the average frequency of positive
slope crossing of the level $u(x) - \bar u =\alpha$. We now
consider how the frequency parameter $\nu_{\alpha}^{+}$ can be
deduced from the underlying probability distributions for $u(x) -
\bar u $. Consider a small length $dl$ of a typical sample
function. Since we are assuming that the process $u(x)-\bar u$ is
a smooth function of $x$, with no sudden ups and downs, if $dl$ is
small enough, the sample can only cross $u(x) - \bar u=\alpha$
with positive slope if $u(x)-\bar u < \alpha$ at the beginning of
the interval location $x$. Furthermore there is a minimum slope at
position $x$ if the level $u(x)- \bar u = \alpha$ is to be crossed
in interval $dl$ depending on the value of $u(x)- \bar u$ at
location $x$. So there will be a positive crossing of $u(x)-\bar u
=\alpha$ in the next space interval $dl$ if, at position $x$,

\begin{equation}
u(x)- \bar u < \alpha \hspace{.6cm} and \hspace {.6cm}
\frac{d(u(x)-\bar u)}{dl}> \frac{\alpha-(u(x) - \bar u) }{dl}.
\end{equation}

%%%%%%%%%%%%%%%%%%%%%%%%%%%%%%%%%%%%%%%%%%%%%%%%%%%%%%%%%%%%%%%%%%%%%%%%%
\begin{figure}
\epsfxsize=9truecm \epsfbox{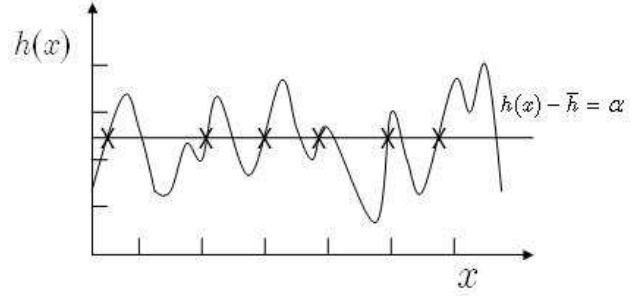} \narrowtext \caption{ positive
slope crossing of the level $ u(x) - \bar u = \alpha$.}
\end{figure}
%%%%%%%%%%%%%%%%%%%%%%%%%%%%%%%%%%%%%%%%%%%%%%%%%%%%%%%%%%%%%%%%%%%%%%%%%%%

Actually what we really mean is that there will be high
probability of a crossing in interval $dl$ if these conditions
are satisfied \cite{s,d}. \\
In order to determine whether the above conditions are satisfied
at any arbitrary location $x$, we must find how the values of $y=
u(x) - \bar u $ and $ y ^{\prime}= \frac{ dy }{dl}$ are
distributed by considering their joint probability density
$p(y,{y}^{\prime})$. Suppose that the level $y=\alpha$ and
interval $dl$ are specified. Then we are only interested in
values of $y < \alpha$ and values of
${y}^{\prime}=(\frac{dy}{dl}) > \frac{\alpha-y}{dl}$, which means
that the region between the lines $y=\alpha$ and ${y}^{\prime}=
\frac{\alpha-y}{dl}$ in the plane ($y,{y}^{\prime}$). Hence the
probability of positive slope crossing of $y=\alpha$ in $dl$ is:
\begin{equation}\label{81}
\int_{0}^{\infty} d{y}^{\prime}\int_{\alpha-{y}^{\prime}dl}^{\alpha}
dy p(y,{ y}^{\prime}).
\end{equation}
When $dl\rightarrow 0$, it is legitimate to put:
\begin{equation}
p(y,{y}^{\prime})=p(y=\alpha,{y}^{\prime}).
\end{equation}
Since at large values of $y$ and ${y}^{\prime}$ the probability
density function approaches zero fast enough, therefore
eq.(\ref{81}) may be written as:
\begin{equation}
\int_{0}^{\infty}
d{y}^{\prime}\int_{\alpha-{y}^{\prime}dl}^{\alpha} dy
p(y=\alpha,{y}^{\prime}),
\end{equation}
in which the integrand is no longer a function of $y$ so that the first
integral is just: $\int_{\alpha-{y}^{\prime}dl}^{\alpha} dy p(y=\alpha,{y}
^{\prime})=p(y=\alpha,{y}^{\prime}){y}^{\prime}dl $,  so that the
probability of slope crossing of $y=\alpha$ in $dl$ is equal to:

\begin{equation}
dl\int_{0}^{\infty}
p(\alpha,{y}^{\prime}){y}^{\prime}d{y}^{\prime},
\end{equation}
in which the term $p(\alpha,{y}^{\prime})$ is the joint probability density $
p(y,{y}^{\prime})$ evaluated at $y=\alpha$.

We have said that the average number of positive slope crossing in
scale $L$ is $\nu_{\alpha}^{+} L$, according to eq.(\ref{3}). The
average number of crossing in interval $dl$ if therefore
$\nu^{+}_{\alpha}dl$. So average number of positive crossing of
$y=\alpha$ in interval $dl$ is equal to the probability of positive
crossing of $y=a$ in $dl$, which is only true because $dl$ is small
and the process $y(x)$ is smooth so that there cannot be more than
one crossing of $y=\alpha$ in space interval $dl$, Therefore we have
$\nu_{\alpha}^{+}dl=dl\int_{0}^{\infty}p(\alpha,{y}^{\prime}){y}
^{\prime}d{y}^{\prime}$, from which we get the following result for
the frequency parameter $\nu_{\alpha}^{+}$ in terms of the joint
probability density function $p(y , {y}^{\prime})$ as follows:
\begin{equation}
\nu_{\alpha}^{+}=\int_{0}^{\infty}p(\alpha,{y}^{\prime}){y}^{\prime}d{y}
^{\prime}.
\end{equation}

In the following section we are going to derive the
$\nu_{\alpha}^{+}$ via the joint PDF of $u(x)-\bar u$ and velocity
gradient. To derive the joint PDF we use the master equation method
\cite{10}. This method enables us to find the $ \nu_{\alpha}^{+}$ in
terms of generating function.
%$Z(\lambda,\mu,x,t) = \langle \exp(-i\lambda (u(x,t))-i\mu
%\omega(x,t))\rangle$, where $\omega(x,t) =
%\frac{\partial}{\partial x} u$.

\section{Frequency of a Definite velocity With Positive Slope for Burgers Equation
Before the Singularity Formation} As mentioned in section I, in
the presence of a random force $\vec{f}(\vec{x}, t)$, the velocity
field of burgers equation in $1+1$ dimensions evolves as:
\begin{eqnarray}
\frac{\partial u}{\partial t} +u\frac{\partial u}{\partial x} =
\nu\frac{\partial^2 u}{\partial x^2}+f(x,t). \label{burg1}
\end{eqnarray}
Differentiating the Burgers equation (eq.\ref{burg1}) respect to
$x$, we have:
\begin{eqnarray}
\frac{\partial \omega}{\partial t } +u\frac{\partial
\omega}{\partial x}+ \omega^2 = \nu \frac{\partial^ 2
\omega}{\partial x^2}+ f_x(x,t),\label{burg2}
\end{eqnarray}
where $\nu \geq 0 $, $\omega=\frac{\partial u}{\partial x}$ and
$f(x,t)$ is a random force, with a Gaussian distribution of mean
zero and second moment given by:
\begin{equation}
\langle f(x,t)f(x^{\prime },t^{\prime })\rangle =2D_0D(x-x^{\prime
})\delta (t-t^{\prime })
\end{equation}
where $D(x)$ is space correlation function and is an even
function of its argument. It has the following form:
\begin{equation}
D(x-x^{\prime })=\frac 1{\sqrt{\pi }\sigma}e^{ -\frac{(x-x^{\prime
})^2}{ \sigma ^2}}
\end{equation}
%\begin{equation}
%D(x-x^{\prime })=\frac 1{\sqrt{\pi }\sigma }\exp
%(-\frac{(x-x^{\prime })^2}{ \sigma ^2})
%\end{equation}
where $\sigma $ is the standard deviation of $D(x-x^{\prime })$.
Typically in the  realistic problem, the correlation of forcing is
considered as smooth function for mimicking the long range
correlation. We regularize the smooth function correlation by a
gaussian function. When the variance $\sigma $ is in order of the
system size, we would expect that the model would represent a long
range character for the forcing. So we should stress that our
calculations are done for finite $\sigma \sim L$, where $L$ is the
system size. Parameters $\nu$ and $D_0$ are describing kinematics
viscosity and the noise strength, respectively. Once trying to
develop the statistical theory of the Burgers equation it becomes
clear that the inter-dependency of the velocity field and velocity
gradient statistics would be taken into account. The very
existence of the non-linear term in the Burgers equation leads to
development of the singularities in a {\it finite time} and in the
inviscid limit i.e. $\nu \rightarrow 0$. So one would distinguish
between different time regimes. Recently it has been shown that
starting from the flat interface the Burgers equation will develop
shock singularity after time scale $t^*$, where $t^*$ depends on
the forcing properties as $t^*\simeq {D_0}^{-1/3} \sigma $
\cite{bah05,10,sha03}. This means that for time scales before
$t^* $ the relaxation contribution tends to zero when $\nu
\rightarrow 0$. In this regime one can observe that the
generating function equation is closed
\cite{bah05,sha03,bahram1,bahram2}. Let us define the generating
function $Z(\lambda, \mu,x,t)$ as:
\begin{eqnarray}
 Z(\lambda,\mu,t)& =& \langle e^{-i\lambda
(u(x,t)-\bar u)-i\mu\omega(x,t)}\rangle =\langle\Theta\rangle.
\end{eqnarray}
 Assuming statistical homogeneity i.e. $ Z_x = 0$ it follows
from equations (\ref{burg1}) and (\ref{burg2}) that $Z$ satisfies
in the following equation:
\begin{eqnarray}
\frac{\partial Z}{\partial t}=-i\lambda\langle\frac{\partial
u}{\partial t}\Theta\rangle-i\mu\langle\frac{\partial^2
u}{\partial t \partial x}\Theta\rangle
\end{eqnarray}
Using the Novikove theorem \cite{bah05} gives:
\begin{equation}
\label{z1} \frac{\partial Z}{\partial t}=i Z_{\mu}-i \mu
Z_{\mu,\mu}-\lambda^2 k(0)Z+\mu^2k_{xx}(0)Z
\label{z1}\end{equation} where $k(x-x^{\prime}) = 2 D_0
D(x-x^{\prime}) $, $ k(0) = \frac{2D_0}{\sqrt{\pi} \sigma}$ and
$k_{xx}(0) =- \frac{4D_0}{\sqrt{\pi } \sigma^3}$. The solution of
eq.(\ref{z1}) by using the separation of variables is as follows:
\begin{eqnarray}
Z(t,\mu,\lambda)&=&\sqrt{\frac{k(0)t}{\pi}}e^{-\lambda^2k(0)t}\times
Z_1(\mu,t)
\end{eqnarray}
where $Z_1(\mu,t)$ satisfies in the following equation:
\begin{eqnarray}
\frac{\partial Z_1}{\partial t}&&=i Z_{1\mu}-i \mu
Z_{1\mu,\mu}-\lambda^2 k(0)Z_1\nonumber\\
&&+[\mu^2k_{xx}(0)+C(t)]Z_1
\end{eqnarray}
$C(t)$ is an arbitrary function which should be determined by
initial conditions.

The joint probability density function of $u$ and $\omega$ can be
obtain by Fourier transform of the generating function:

\begin{equation}
P(u,\omega,t)=\frac{1}{2\pi}\int d\lambda d\mu e^{i\lambda (u-\bar
u)+i\mu \omega}Z(\lambda,\mu,t),
\end{equation}
so by Fourier transforming of the eq.(\ref{z1}) we get the
Fokker-Planck equation as:

\begin{equation}
\frac{\partial}{\partial t}P= 3\omega P+\omega^2\frac{\partial
P}{\partial \omega}+k(0)\frac{\partial^2 P}{\partial u^2
}-k_{xx}(0)\frac{\partial^2 P}{\partial \omega^2}.
\end{equation}

The solution of the above equation can be separated as $
P(u,\omega,t)=p_{1}(u,t)p_{2}(\omega,t)$ (for motivation
see\cite{seri}). Using the initial conditions $
P_{1}(u,0)=\delta(u)$ and $P_{2}(\omega,0)=\delta(\omega)$  it can
be shown that:
\begin{eqnarray}\label{propability}
P(u,\omega,&t&) = \frac{1}{\sqrt{4\pi
k(0)t}}e^{-\frac{u^2}{4k(0)t}}\times p_{2}(\omega,t)
\end{eqnarray}
where $p_{2}(\omega,t)$ is a solution of the following equation:
\begin{equation}
\frac{\partial}{\partial t}p_{2}= +\omega^2\frac{\partial
p_{2}}{\partial \omega}-k_{xx}(0)\frac{\partial^2 p_{2}}{\partial
\omega^2}+[3\omega+G(t)] p_{2}(\omega,t) \label{p2}
\end{equation}
$G(t)$ is an arbitrary function which should be determined by
initial conditions. Up to now we obtained that, for every time
scale, the level cross $\nu_{\alpha}^+$ has a Gaussian behavior in
terms of $\alpha$, now for determination of time dependence of
$\nu_{\alpha}^+$ we have to determine $p_{2}(\omega,t)$. Since the
eq.(\ref{p2}) is so complex, we solve it using the numerical
methods \cite{numeric}. The frequency of repeating a definite
velocity field $(u(x)- \bar u = \alpha)$ with positive slope can
be calculated as $ \nu_{\alpha}^{+}=\int_{0}^{\infty}\omega
P(\alpha,\omega,t)d\omega$. Fig.2 shows $\nu_{\alpha}^{+}$ for
various time scales before creation of singularity. To derive the
$N_{tot}^{+}$ let us express $ N_{tot}^{+}$ as:
\begin{equation}\label{n}
N_{tot}^{+}=\int_{-\infty }^{+\infty }d\alpha\int^{\infty}_{0}
\omega P(\alpha,\omega,t)d \omega
\end{equation}
Using the numerical integration of eq.(\ref{n}) one finds
$N_{tot}^{+} \sim t^{\beta}$ where $\beta=0.50\pm0.01$, In fig.(3)
we plot the $N^+_{tot}$ as a function of $t$.

%%%%%%%%%%%%%%%%%%%%%%%%%%%%%%%%%%%%%%%%%%%%%%%%%%%%%%%%%%%%%%%%%%%%%%%%%
\begin{figure}
\epsfxsize=9.7truecm \epsfbox{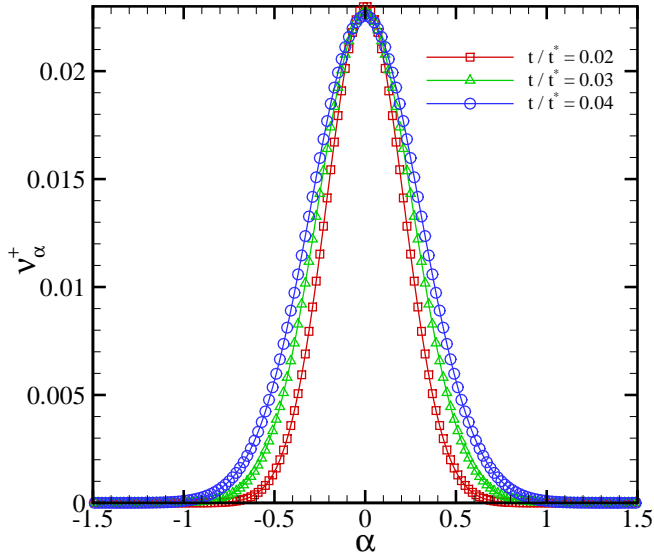} \narrowtext \caption{ plot
of $\nu_{\alpha}^{+}$ vs $\alpha$ for the Burgers equation in the
strong coupling and before the creation of singularity for time
scale $t/t* = 0.02, 0.03$ and $0.04$. }
\end{figure}
%%%%%%%%%%%%%%%%%%%%%%%%%%%%%%%%%%%%%%%%%%%%%%%%%%%%%%%%%%%%%%%%%%%%%%%%%%%

%%%%%%%%%%%%%%%%%%%%%%%%%%%%%%%%%%%%%%%%%%%%%%%%%%%%%%%%%%%%%%%%%%%%%%%%%
\begin{figure}
\epsfxsize=9.7truecm \epsfbox{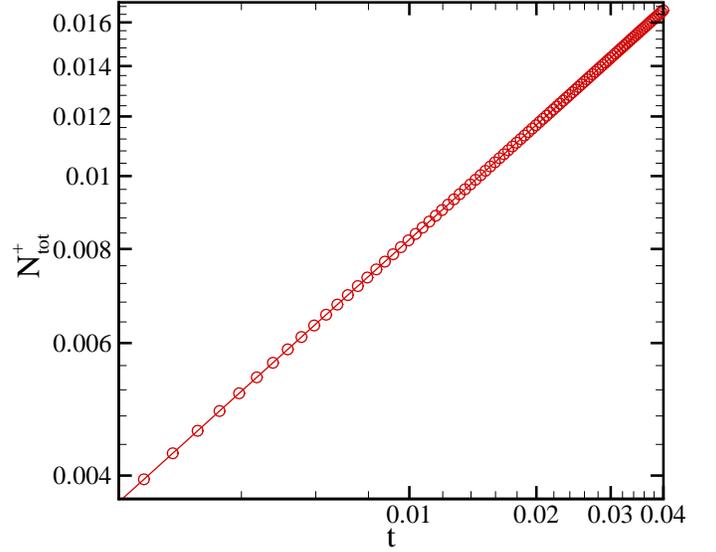} \narrowtext \caption{
log-log plot of $N_{tot}^{+}$ vs $t$ for the Burgers equation in
the strong coupling before the creation of singularity in the
velocity field. }
\end{figure}
%%%%%%%%%%%%%%%%%%%%%%%%%%%%%%%%%%%%%%%%%%%%%%%%%%%%%%%%%%%%%%%%%%%%%%%%%%%
\section{conclusion}

 We obtained some results in the problems of Burgers
equation in 1$+$1 dimensions with a Gaussian forcing which is
white in time and  Gaussian correlated in space, typically in the
physical case. We determined the average frequency of crossing
i.e. $\nu _\alpha ^{+}$ of observing of the definite value for
velocity field $u-\bar{u}=\alpha $, from which one can find the
averaged number of crossing the given velocity field in a sample
with size L. The integral representation of $ \nu _\alpha ^{+}$
was given for the Burgers equation in the inviscid limit before
the creation of singularity and  it was shown that the velocity
dependence of the $ \nu _\alpha ^{+}$ is Gaussian. We apply the
quantity $ N_{tot}^{+}=\int_{-\infty }^{+\infty }\nu _\alpha
^{+}d\alpha $, which measures the total number of positive
crossing of velocity and show that for the Burgers equation in the
inviscid limit and before the creation of singularities
$N_{tot}^{+}$ scales as $t^{1/2}$.

 {\bf Acknowledgment} We would like to thank M R Rahimi Tabar  and  M Fazeli for useful comments. My special thanks to
H Abdollahi for discussions about numerical calculations. This paper
is dedicated to Dr Somaieh Abdolahi.


\begin{references}

\bibitem{cos} S.N. Gurbatov, A.I. Saichev and S.F. Shandarin, The
large-scale structure of the Universe in the frame of the model
equation of non-linear diffusion, Monthly Notices of the Royal
Astronomical Society 236:385–402 (1989).

\bibitem{traf} D. Chowdhury, L. Santen and A. Schadschneider, Statistical
physics of vehicular traffic, Phys. Rep. 329:199–329 (2000).

\bibitem{med} J.-P. Bouchaud, M. M´ezard and G. Parisi, Scaling
and intermittency in Burgers turbulence, Phys. Rev. E
52:3656–3674 (2000).

\bibitem{KPZ} M. Kardar, G. Parisi and Y.-C. Zhang, Dynamic scaling of
growing interfaces, Phys. Rev. Lett. 56:889–892 (1986).

\bibitem{ya} Ya.G. Sinai, Two results concerning asymptotic behavior of the
solutions of the Burgers equation with force, J. Stat. Phys.
64:1–12 (1991).

\bibitem{ch} A. Chekhlov and V. Yakhot, Kolomogorov turbulence in a
random-force-driven Burgers equation, Phys. Rev. E 51:R2739–R2742
(1995).

\bibitem{w1} W. E, K. Khanin, A. Mazel and Ya.G. Sinai, Probability
distribution functions for the random forced Burgers equation,
Phys. Rev. Lett. 78:1904–1907 (1997).

\bibitem{w2} W. E, K. Khanin, A. Mazel and Ya.G. Sinai, Invariant
measures for Burgers equation with stochastic forcing, Ann. Math.
151:877–960 (2000).


\bibitem{r} R. Iturriaga and K. Khanin, Burgers turbulence and random lagrangian systems. commun. math. phys. {\bf 232}, 3, 377 (2003).


\bibitem{j} J. Bec, R. Iturriaga and K. Khanin, Topological shocks in
Burgers turbulence, Phys. Rev. Lett. 89:024501 (2002).

\bibitem{bah05} A. Bahraminasab, M. Sadegh Movahed, S. D.
Nassiri and A. A. Masoudi, Journal of Statistical Physics, Vol. 124,
No. 6, September 2006.


\bibitem{bara}  A.-L. Barabasi and H. E. Stanley, ''Fractal Concepts in Surface
Growth'' (Cambridge University Press, New York, 1995).

\bibitem{ris} H. Risken, {\it The Fokker-Planck Equation}, (Springer,
Berlin), 1984.

\bibitem{s}  S.O. Rice '' Mathematical Analysis of Random Noise'' Bell
System Tech. J. Vol. 23, (1944), 282; Vol. 24 (1945), 46

\bibitem{d}  D.E. Newland '' An Introduction to Random Vibration, Spectral
and Wavelet Analysis'' ( Longman, Harlow and Wiley, New York,
1993 )
\bibitem{10}  A. A. Masoudi, F. Shahbazi, J. Davoudi and M. Reza Rahimi
Tabar, Phys. Rev. E {\bf 65}, 026132(2002).


\bibitem{sha03} F. Shahbazi, S. Sobhanian, M.R. Rahimi Tabar, S. Khorram, G.R. Frootan,
and H. Zahed, J. phys. A, {\bf 36}, 2517 (2003)


\bibitem{bahram1} A. Bahraminasab, S. M. A. Tabei, A.A.
Masoudi, F. Shahbazi and M. Reza Rahimi Tabar
  Journal of
Statistical Physics {\bf116},  1521 (2004).

\bibitem{bahram2} S. M. A. Tabei, A. Bahraminasab,
A.A. Masoudi, S. S. Mousavi and M. Reza Rahimi Tabar
 Phys. Rev. E {\bf 70}, 031101 (2004).
 \bibitem{seri} Sreenivasan K R, Prabhu A and Narasimha R, J. Fluid Mech. {\bf 137}
 251 (1983).

\bibitem{numeric} J. H. Mathews, {\it Numerical methods for mathematics and science and
engineering}, Prentice Hall College Div; 2 edition (1992).

\end{references}
\end{document}